\documentclass[12pt,reqno]{article}

\usepackage{mathtools}
\usepackage{dirtytalk}
\usepackage{amsmath}
\usepackage{amsfonts}
\usepackage{amssymb}
\usepackage{amsxtra}
\usepackage{amsthm}
\usepackage{graphicx}
\usepackage{caption}
\usepackage{hyperref}
\usepackage{ushort}
\usepackage{color}
\usepackage[parsep]{collref}
\hypersetup{linktocpage=true}
\usepackage{enumitem}
\usepackage{titlesec}
\usepackage{cite}
\usepackage{slashed}
\usepackage{braket}
\usepackage{flexisym}
\usepackage{xcolor}
\usepackage{dsfont}
\usepackage{titlesec}
\usepackage{romannum}
\usepackage{caption}
\usepackage{subcaption}
\usepackage{mathrsfs}
\usepackage[bottom]{footmisc}
\usepackage[bbgreekl]{mathbbol}
\usepackage{bbold}

\newtheorem{theorem}{Theorem}[section]

\DeclareSymbolFontAlphabet{\mathbb}{AMSb}
\DeclareSymbolFontAlphabet{\mathbbl}{bbold}
\def\ie{{\it i.e.\ }}

\newcommand\numberthis{\addtocounter{equation}{1}\tag{\theequation}}

\newcommand{\mb}{\mathbb}
\newcommand{\mc}{\mathcal}

\newcommand{\ov}{\overline}
\newcommand{\ph}{\phantom}

\DeclareMathOperator{\sech}{sech}

\DeclareMathOperator{\vol}{vol}

\DeclareMathOperator{\diag}{diag}

\newcommand{\imperial}{\it 
The Blackett Laboratory, Imperial College London\\
Prince Consort Road, London, SW7 2AZ
}
\newcommand{\auth}{Rahim Leung\,\footnote{\,rahim.leung14@imperial.ac.uk} and K.S. Stelle\,\footnote{\,k.stelle@imperial.ac.uk}}

\let\oldabstract\abstract
\let\oldendabstract\endabstract
\makeatletter
\renewenvironment{abstract}
{%
               {\list{}{\addtolength{\leftmargin}{1em} 
                        \listparindent 1.5em%
                        \itemindent    \listparindent%
                        \rightmargin   \leftmargin%
                        \parsep        \z@ \@plus\p@}%
                \item\relax}%
               {\endlist}%
\oldabstract}
{\oldendabstract}
\makeatother

\numberwithin{equation}{section}

\let\oldsection\section
\renewcommand{\section}{\renewcommand{\theequation}{\thesection.\arabic{equation}}\oldsection}

\titlespacing*{\section}
{0pt}{5.5ex plus 1ex minus .2ex}{4.3ex plus .2ex}
\titlespacing*{\subsection}
{0pt}{5.5ex plus 1ex minus .2ex}{4.3ex plus .2ex}

\begin{document}
\setcounter{page}{0}
\thispagestyle{empty}
\begin{flushright}
\hfill{
Imperial/TP/2023/KS/01}\\
\end{flushright} 
\vspace{15pt}

\begin{center}  

{\Large {\bf Localised Gravity and Resolved Braneworlds}}   

\vspace{15pt}

\auth

\vspace{7pt}
\imperial

\end{center} 

\begin{center}
{\large\bf Dedicated to the Memory of Stanley Deser}
\end{center}

\begin{abstract}

Deriving an effective massless field theory for fluctuations about a braneworld spacetime requires analysis of the transverse-space-wavefunction's second-order differential equation. There can be two strikingly different types of effective theory. For a supersymmetric braneworld, one involves a technically consistent embedding of a supergravity theory on the worldvolume; the other can produce, in certain situations, a genuine localisation of gravity near the worldvolume but not via a technically consistent embedding. So, in the latter situation, the theory's dynamics remains higher-dimensional but there can still be a lower-dimensional effective-theory interpretation of the dynamics at low worldvolume momenta / large worldvolume distances.

This paper examines the conditions for such a gravity localisation to be possible. Localising gravity about braneworld spacetimes requires finding solutions to transverse-space self-adjoint Sturm-Liouville problems admitting a normalisable zero mode in the noncompact transverse space. This in turn requires analysis of Sturm-Liouville problems with radial singular endpoints following a formalism originating in the work of Hermann Weyl. Examples of such gravity-localising braneworld systems are found and analysed in this formalism with underlying ``skeleton'' braneworlds of Salam-Sezgin, resolved D3-brane and Randall-Sundrum II types.

\end{abstract}

\vfill\leftline{}\vfill
\pagebreak

\tableofcontents
\addtocontents{toc}{\protect\setcounter{tocdepth}{2}}
\pagenumbering{arabic}
\setcounter{page}{1}
\setcounter{footnote}{0}

\section{Introduction: Brane dynamics taxonomy}

That superstring theories are about more than just perturbative strings has become clear over the past two decades. The wide array of brane solutions of the corresponding effective supergravity theories fills out a solitonic sector which also carries representations of the associated duality symmetries in the originating and reduced lower spacetime dimensions. Those dualities also are the key links between the variants of perturbatively quantisable superstring theories. The descent through effective theory spacetime dimensions has mostly been considered in the context of dimensional compactification of the extra dimensions transverse to a given lower dimensional spacetime.

Compactification of the transverse dimensions is not the only possibility, however. The array of supergravity brane solution worldvolumes presents another kind of locale in which to situate lower-dimensional physics. There is a basic difference, however: unless they are ``stacked'' \cite{Lu:1996mg} so as to generate compact dimensions transverse to their worldvolumes, brane transverse spaces are generically noncompact. Dimensional reduction on noncompact spaces in turn generically runs into a problem identified by Hull and Warner \cite{Hull:1988jw}, that the transverse wave-equation spectrum is generically a continuum running down to zero eigenvalue, leading to a vanishing Newton constant for coupling the lower dimensional gravity to other fields. Generically does not mean universally, however, as was shown by Randall and Sundrum \cite{Randall:1999vf} in a construction based on sliced and reflected AdS subspaces.

Another situation where noncompact transverse spaces can avoid the problem of a vanishing Newton constant is where Newton's constant is not fundamentally meaningful, owing to the existence (at least at the classical level in effective theory) of a ``trombone'' symmetry \cite{Cremmer:1997xj} such as that in pure vacuum Einstein theory. Pure Einstein theory without external field-theory sources has, up to bothersome issues about GR delta-function singularities \cite{Arnowitt:1960zzb}, black hole solutions such as the Schwarzschild solution. The black-hole orbits of pure GR's  trombone symmetry are parametrised by the mass of the black hole. This is also the situation for pure-gravity worldvolume dynamics obtained by replacing the flat brane worldvolume metric by a Ricci-flat metric \cite{Brecher:1999xf}, producing a consistent truncation of the higher dimensional originating theory down to pure GR on the worldvolume. Moreover, for an underlying ``skeleton'' brane with unbroken supersymmetry, this sort of consistent lower-dimensional theory embedding can extend to a full supergravity theory corresponding to the skeleton unbroken supersymmetry \cite{Leung:2022nhy}. In the terminology of Ref.\ \cite{Erickson:2021psj}, such a consistent-truncation effective-theory origin involves Type I boundary conditions for the transverse-space eigenfunctions as one approaches the skeleton-brane worldvolume.

For transverse space differential equations of second order in derivatives, however, there will be a variety of boundary condition choices as one approaches a brane worldvolume. Type III boundary conditions, when available, can produce a different type of reduced dimension effective theory -- not involving a consistent truncation, but with a genuine dynamical localisation of gravity and companion fields into the region near the brane worldvolume. An initial example \cite{Crampton:2014hia} of such Type III localisation involves the lift to $D=10$ Type IIA supergravity (or further up to $D=11$ M-Theory) of the $D=6$ Salam-Sezgin vacuum (called the SS-CGP solution in the following) \cite{Cvetic:2003xr,Crampton:2014hia}, which involves a hyperbolic ${\cal H}^{(2,2)}$ transverse-space metric. That transverse-space geometry allows for a dynamical localisation of gravity near the lower-dimensional worldvolume owing to a special feature of the transverse-space wave equation. When transformed into a Schr\"odinger equation form, it has a P\"oschl-Teller potential for which the spectrum has a single $L^2$ normalisable bound state separated from the edge of the continuous spectrum by a mass gap. It is the existence of the $L^2$ normalisable transverse state that gives rise to gravity localisation near the lower-dimensional worldvolume. The required boundary conditions on the transverse spectrum is of Robin form for this Type III system, distinct from the Dirichlet-skeleton conditions of a Type I reduction.\footnote{Distinctions between boundary condition choices are also related to distinctions between possible source terms at transverse-space endpoints. Consideration of source terms have been treated in Ref.\ \cite{Crampton:2014hia} and also more recently in Ref.\ \cite{DeLuca:2023kjj}.}

In this paper, we begin in Section \ref{sec:D3res} with an initial example that allows for such a choice between either a Type I consistent supergravity embedding or a Type III localisation, involving a resolved D3 brane with an Eguchi-Hanson transverse geometry. This is related, via a sequence of dualities, to the SS-CGP skeleton of Ref.\ \cite{Crampton:2014hia}. 

This is followed in Section \ref{sec:Localisedgravitons} with a review of  the formalism for localised graviton fluctuations and the transverse-space wave equation about a ``skeleton'' quiescent brane, particularly for resolved branes.

In Section \ref{sec:SLT}, we set the stage for  a general discussion of localisation possibilities based on Sturm-Liouville theory for systems with singular endpoints, originating in the work of Hermann Weyl \cite{Weyl:1910} and fully developed in the work of Anton Zettl and collaborators, as presented in Ref.\ \cite{Zettl:2005}. The heart of the matter is a characterisation of the available boundary conditions as one approaches the lower-dimensional worldvolume and at infinity. 

We then apply this analysis in Section \ref{sec:examples} to five examples where boundary conditions giving gravity localisation are successfully found and also one example where localisation is not found. We end with a conjecture on the relevance of a two-dimensional subspace of the transverse space and the attendant logarithmic structure of the normalisable transverse zero mode.

After the Conclusion in Section \ref{sec:Concl}, we give four technical Appendices.

\section{Resolved D3-branes on Eguchi-Hanson space: consistently embedded worldvolume supergravity versus genuine localisations}\label{sec:D3res}

We'll consider here, a particular system of resolved D3-branes by transgression, the general formalism of which is discussed in Ref.\ \cite{Cvetic:2000mh}. Here, the RR and NSNS three-form fluxes are present and depend on the six-dimensional space transverse to the worldvolume of the branes. Such a configuration is also given the name of fractional D3-branes, and are holographically dual to $\mc{N}=1$ super Yang-Mills in four dimensions \cite{Klebanov:2000nc,Klebanov:2000hb,Grana:2000jj,Gubser:2000vg,PandoZayas:2000ctr}. 
Our conventions for M-theory, Type IIA and Type IIB supergravities are given in Appendices \ref{Mtheory} and \ref{IIBeoms}.

The resolved D3-branes that we consider are geometrically a warped product $\mb{R}^{1,3}\times_W( M_4\times T^2)$, where $M_4$ is the Eguchi-Hanson space. To obtain such a solution, we will follow the recipe of \cite{Cvetic:2000mh} and propose the ansatz\footnote{In the language of \cite{Cvetic:2000mh}, we have $\hat G_{(3)} = L_{(3)}$, where $L_{(3)}$ is a complex self-dual three-form satisfying ${\ast}_6L_{(3)} = iL_{(3)}$, where ${\ast}_6$ is the Hodge dual with respect to the Ricci-flat metric on $M_4\times T^2$.}
\begin{equation}\label{D3EH}
\begin{split}
&d\hat s^2 = H^{-1/2}\eta_{\mu\nu}dx^\mu dx^\nu + H^{1/2}(ds^2_{EH} + dzd\ov z) \,,\quad \hat\Phi = 0 \,,\quad \hat C_0 = 0 \,,\\
&\hat G_{(3)} = \mc{F}_{(2)}\wedge d\ov z\,,\quad \hat F_{(5)} = H^{-2}\vol_4\wedge dH - \frac{i}{2}{\ast}_{EH}dH \wedge dz\wedge d\ov z \,.
\end{split}
\end{equation}
Here, $\vol_4$ is the volume form on $\mb{R}^{1,3}$, $z$ is the complex coordinate on $T^2$, the Ricci-flat metric on the Eguchi-Hanson space is
\begin{equation}\label{EHmetric}
ds^2_{EH} = \cosh2\rho\left(d\rho^2 + \frac{1}{4}\tanh^22\rho\,(d\chi+\cos\theta\,d\varphi)^2 + \frac{1}{4}(d\theta^2 + \sin^2\theta\,d\varphi^2)\right) \,,
\end{equation}
with $\rho\geq0$, $\chi\in[0,2\pi)$, $\theta\in[0,\pi]$ and $\varphi\in[0,2\pi)$; ${\ast}_{EH}$ is the Hodge dual with respect to $ds^2_{EH}$, $H$ is a function on the Eguchi-Hanson space, and $\mc{F}_{(2)}$ is the unique (up to normalisation), self-dual harmonic two-form on the Eguchi-Hanson space given by
\begin{equation}\label{EHharmonic2form}
\mc{F}_{(2)} = d\mc{A}_{(1)}\,,\quad \mc{A}_{(1)} = \frac{1}{2}\sech2\rho\,(d\chi+\cos\theta\,d\varphi)\,.
\end{equation}
All of the equations of motion are satisfied provided that $H$ is a solution to the Poisson equation
\begin{equation}\label{poissonresolvedD3}
d{\ast}_{EH}dH = -\mc{F}_{(2)}\wedge{\ast}_{EH}\mc{F}_{(2)} \,.
\end{equation}
Assuming spherical symmetry, $H = H(\rho)$, the explicit solution is given by
\begin{equation}\label{sourcedharmonicEH}
H = \sech2\rho - k\log\tanh\rho \,,
\end{equation}
where the logarithmic term is the homogeneous solution of the Poisson equation \eqref{poissonresolvedD3}. For $H$ to be positive definite for all $\rho$, we will take $k \geq 0$. When $k= 0$, the solution \eqref{D3EH} is smooth for all values of $\rho$, and interpolates between the Ricci-flat vacuum $\mb{R}^{1,3}\times M_4\times T^2$ at small $\rho$ and the warped $\mb{R}^{1,3}\times\mb{R}\text{P}^3\times \mb{H}^3$ solution at large $\rho$, where the $\mb{R}\text{P}^3$ is the constant $\rho$ slice of the Eguchi-Hanson space. When $k>0$, the large $\rho$ asymptotic structure remains the same, but there is a curvature singularity at $\rho=0$.

Our solution \eqref{D3EH} is a cousin of the Type IIA solution of \cite{Cvetic:2003xr} which describes the ten-dimensional embedding of the $\mb{R}^{1,3}\times S^2$ Salam-Sezgin vacuum with an NS5-brane inclusion, where the NS5-brane has a tension proportional to the parameter $k$. This solution, which we will call the CGP-SS solution, is 
\begin{equation}\label{SSNS}
\begin{split}
&d\tilde s^2 = H^{-1/4}\left(\eta_{\mu\nu}dx^\mu dx^\nu + dy^2 + (d\psi + \mc{A}_{(1)})^2\right) + H^{3/4}ds^2_{EH} \,,  \\
&e^{2\tilde\phi} = \sech2\rho\,,\quad \tilde B_{(2)} = \frac{1}{4}(1+k)\cos\theta\,d\chi\wedge d\varphi - \mc{A}_{(1)}\wedge d\psi \,, 
\end{split}
\end{equation} 
where $H$ is given by \eqref{sourcedharmonicEH}, the Salam-Sezgin gauge parameter $\ov g$ is set to $1/\sqrt{2}$ for convenience, and we use tildes to distinguish the Type IIA fields from the Type IIB ones.

Although the geometric structure of our resolved D3 solution and the CGP-SS solution are different, they are related by a series of dualities. In terms of branes, this duality map is 
\begin{equation}\label{dualitymap}
\text{NS5}\underbrace{\to}_{\text{M-theory/IIA}}\text{D4}\underbrace{\to}_{\text{T-duality}}\text{D3} \,.
\end{equation}
To construct this explicitly, we first lift the CGP-SS solution to M-theory on a circle with coordinate $\zeta$, then reduce back to Type IIA on the worldvolume circle $y$. The resulting Type IIA solution describes a resolved D4-brane, with
\begin{equation}
\begin{split}
&d\tilde s^2_{str} = H^{-1/2}\left(\eta_{\mu\nu}dx^\mu dx^\nu +\left(d\psi + \mc{A}_{(1)}\right)^2\right) + H^{1/2}\left(ds^2_{EH} + d\zeta^2\right) \,,\\
&e^{2\tilde\phi} = H^{-1/2} \,,\quad \tilde F_{(4)} = (2(1+k)\vol(\mb{R}\text{P}^3) - \mc{F}_{(2)}\wedge d\psi)\wedge d\zeta\,,
\end{split}
\end{equation}
where 
\begin{equation}
\vol(\mb{R}\text{P}^3) = \frac{1}{8}\sin\theta\,d\chi\wedge d\theta\wedge d\varphi 
\end{equation}
is the volume form of the asymptotic $\mb{R}\text{P}^3$ at infinity of the Eguchi-Hanson space. We have expressed the solution in string frame, $d\tilde s^2_{str}= e^{\tilde\phi/2}d\tilde s^2_{Ein}$, which is more convenient for T-dualisation. Applying the Bucher rules for Abelian T-duality on the circle $\psi$, which are detailed in Appendix \ref{Bucher}, we obtain the Type IIB solution
\begin{equation}\label{SSD3}
\begin{split}
&d\hat s^2 = H^{-1/2}\eta_{\mu\nu}dx^\mu dx^\nu + H^{1/2}\left(ds^2_{EH} +d\zeta^2 + d\xi^2 \right) \,,\quad \hat\Phi = 0\,,\quad \hat C_0 = 0\,, \\
&\hat H_{(3)}= -\mc{F}_{(2)}\wedge d\xi \,,\quad \hat F_{(3)} = \mc{F}_{(2)}\wedge d\zeta\,,\\
&\hat F_{(5)} = 2(k+\tanh^22\rho)(1+\hat{\ast})\vol(\mb{R}\text{P}^3)\wedge d\zeta\wedge d\xi \,,
\end{split}
\end{equation}
where $\xi$ is the T-dual coordinate of $\psi$. Identifying the complex coordinate $z = \zeta + i\xi$, we find that \eqref{SSD3} is exactly our resolved D3-brane solution \eqref{D3EH}. Since the original CGP-SS solution preserves eight rigid supercharges with a Killing spinor that is independent of the T-dualised coordinate $\psi$ \cite{Crampton:2014hia}, our resolved D3-brane solution \eqref{D3EH} also preserves eight rigid supercharges. 

A straightforward extension of this is the construction of the Type IIB T-dual of the Type IIA embedding of the full six-dimensional Salam-Sezgin theory. This is done in Appendix \ref{Salam-Sezgin T-dual}. The result is an embedding of Type IIB supergravity on a four-dimensional K\"ahler manifold that is a product $M_2\times T^2$, where the metric on $M_2$ is the induced metric on the Eguchi-Hanson space at constant values on the two-sphere. Properties of this four-manifold, including a construction of its K\"ahler structure, are explored in Appendix \ref{Salam-Sezgin T-dual}.

The fact that \eqref{D3EH} preserves eight rigid supercharges suggests that we can embed the four-dimensional $\mc{N}=2$ supergravity into its worldvolume, just as for the usual half-supersymmetric D3-branes. The bosonic sector of the $\mc{N}=2$ theory is the Einstein-Maxwell theory with Lagrangian
\begin{equation}
\mc{L}_4 = R{\ast}_41 - \frac{1}{2}F_{(2)}\wedge{\ast}_4F_{(2)} \,.
\end{equation}
Using similar arguments to \cite{Leung:2022nhy}, one may propose the consistent-truncation ansatz 
\begin{equation}\label{D3EHN=2sugra}
\begin{split}
&d\hat s^2 = H^{-1/2}g_{\mu\nu}dx^\mu dx^\nu + H^{1/2}(ds^2_{EH} + dzd\ov z) \,,\quad \hat\Phi = 0 \,,\quad \hat C_0 = 0 \,,\\
&\hat G_{(3)} = \mc{F}_{(2)}\wedge d\ov z + \frac{1}{2}(F_{(2)}+i{\ast}_4F_{(2)})\wedge dz\,,\quad \\
&\hat F_{(5)} = H^{-2}\vol_4\wedge dH - \frac{i}{2}{\ast}_{EH}dH \wedge dz\wedge d\ov z \,,
\end{split}
\end{equation}
where $\vol_4$ is the volume form associated to the four-dimensional metric $g_{\mu\nu}$, and ${\ast}_4$ is the Hodge dual with respect to $g_{\mu\nu}$. We find that the Type IIB equations of motion and Bianchi identities are satisfied provided that $(g_{\mu\nu}, F_{(2)})$ satisfy the equations of motion and the Bianchi identity of $\mc{N}=2$ supergravity in four dimensions. We can use this to obtain an embedding of the $\mc{N}=2$ supergravity on the worldvolume of the CGP-SS solution by reversing the duality map \eqref{dualitymap}.

In the language of \cite{Erickson:2021psj}, the embedding \eqref{D3EHN=2sugra} is a Type I embedding of Type IIB supergravity to the $\mc{N}=2$ supergravity in four dimensions. But there is another possibility. One striking property of the CGP-SS solution is that, although the Eguchi-Hanson space is non-compact, transverse-traceless graviton perturbations on the NS5-brane worldvolume having a non-trivial dependence on the Eguchi-Hanson radial coordinate $\rho$ may instead be genuinely localised \cite{Crampton:2014hia,Erickson:2021psj}. In the language of Ref.\ \cite{Erickson:2021psj}, this gives a Type III embedding. As we will now show, this is also a property of our resolved D3-brane solution \eqref{D3EH}. Consider a transverse-traceless perturbation of the Minkowski worldvolume that depends both on the worldvolume coordinates as well as the Eguchi-Hanson radius $\rho$,
\begin{equation}
\eta_{\mu\nu}\mapsto \eta_{\mu\nu} + \mc{H}_{\mu\nu}(x,\rho) \,,\quad \partial^\mu\mc{H}_{\mu\nu}= 0 \,,\quad \eta^{\mu\nu}\mc{H}_{\mu\nu} = 0 \,.
\end{equation}
From the general results of \cite{Bachas:2011xa}, the equations of motion to first-order in perturbations are satisfied provided that
\begin{equation}
(\Box_4 + H^{-1}\Box_{EH})\mc{H}_{\mu\nu}(x,\rho) =0 \,,\quad H = \sech2\rho - k\log\tanh\rho \,,
\end{equation}
where $\Box_4$ and $\Box_{EH}$ are the respectively the Laplacians on $\mb{R}^{1,3}$ and Eguchi-Hanson space. For functions that only depend on $\rho$, 
\begin{equation}
\Box_{EH}f(\rho) = \frac{1}{\sinh2\rho\cosh2\rho}\frac{d}{d\rho}\left(\sinh2\rho\frac{d}{d\rho}f(\rho)\right) \,.
\end{equation}
The graviton spectrum can be identified by performing a decomposition of $\mc{H}_{\mu\nu}$ in terms of the complete set of eigenfunctions of the operator $H^{-1}\Box_{EH}$,
\begin{equation}
\mc{H}_{\mu\nu}(x,\rho) = h^{(\lambda)}_{\mu\nu}(x)\xi_{(\lambda)}(\rho) \,,\quad H^{-1}\Box_{EH}\xi_{(\lambda)} = -\lambda^2\xi_{(\lambda)} \,,
\end{equation}
where $\lambda\in\mb{R}$, and there is a sum/integral over the $\lambda$ index. The mode corresponding to $\lambda = 0$ is a massless graviton, and the modes with $\lambda \neq 0$ are massive with mass $|\lambda|$. Localisation is achieved if the zero eigenfunction $\xi_0(\rho)$ is square-integrable with respect to the induced measure $\mu = H\mu_{EH}$, where $\mu_{EH} = \sinh2\rho\cosh2\rho$ is the $\rho$-dependent part of the measure on the Eguchi-Hanson space. If there is a mass gap separating the zero mode from the massive modes, corrections to the effective four-dimensional gravitational potential are of Yukawa form, otherwise they are polynomial. The localisation of gravity on the brane worldvolume thus depends solely on the spectrum of the operator $H^{-1}\Box_{EH}$. Fortunately, this is the same operator as that considered in \cite{Crampton:2014hia,Erickson:2021psj}. By imposing the Robin boundary condition at $\rho = 0$
\begin{equation}
((\sinh2\rho)\log\tanh\rho\,\partial_\rho - 2)\xi_{(\lambda)}(\rho)\big\rvert_{\rho = 0} = 0 \,,
\end{equation}
and requiring normalisability, in Refs \cite{Crampton:2014hia,Erickson:2021psj} it was shown that the spectrum of the operator consists of a single zero mode bound state 
\begin{equation}
\xi_{(0)}(\rho) = \frac{1}{\pi}\left(\frac{24}{2+3k}\right)^{1/2}\log\tanh\rho\,,
\end{equation}
where the normalisation constant is defined so that
\begin{equation}
\int_0^\infty d\rho\,\mu\,\xi_{(0)}^2 = 1 \,,
\end{equation}
and a continuum of massive mode scattering states with eigenvalues $\lambda^2 > 1+k$. This gives a mass gap of $\delta = 1+k$.

In the  Type III perspective of the effective field theory for the graviton modes, it is inconsistent to truncate away the massive gravitons. This is because the massive gravitons couple to the massless graviton at cubic order in the action with coefficients proportional to the overlap integrals of $\xi_{(0)}^2\xi^{}_{(\lambda)}$ or $\xi_{(0)}\xi_{(\lambda)}\xi_{(\lambda')}$. Neither of these integrals vanish, and the same situation occurs also at higher orders in the action. However, due to the non-zero mass gap $\delta$, the effective theory obtained by integrating out the massive modes is a valid theory up to the energy scale defined by $\delta$. Such effective theories of massless modes supported by non-constant zero eigenfunctions were studied in \cite{Erickson:2020oda,Erickson:2022qhv}. A particular feature of these theories is that a delayed, covert form of spontaneous symmetry breaking can occur, revealing itself only at fourth order in the action.\footnote{Symmetry breaking at fourth order in fields is characteristic of situations where field redefinitions are incapable of restoring a symmetry that is apparent at lower order \cite{Deser:2019yig}.} The interaction coefficients at fourth order and higher do not to have the values one expects for a generally-covariant theory in an order-by-order field expansion \cite{Arnowitt:1962hi,Weinberg:1965rz}. A quick way to see this is to note that the $n^{\text{th}}$-order terms in the effective action involving just the massless mode are proportional to the integral
\begin{equation}
I_n = \int_0^\infty d\rho\,\mu\,\xi_{(0)}^n \,.
\end{equation}
In the perturbative formulation of gravity, the fourth-order coefficient is the square of the third-order coefficient. However, an explicit calculation shows that $I_4 \neq (I_3)^2$. This does not mean that the gauge symmetry is broken. As shown in \cite{Erickson:2020oda}, the gauge symmetry is preserved due to the existence of a field with a Stueckelberg-like transformation that only appears when we consider the full, ten-dimensional gravity perturbation $\hat g_{MN} \mapsto \hat g_{MN} + \hat{H}_{MN}$, and not just the worldvolume, transverse-traceless part. This Stueckelberg-like field was shown in \cite{Erickson:2022qhv} to be a \say{phantom} mode -- it only contributes to the energy of the effective field theory at interacting level. 

\section{Localised worldvolume gravitons on resolved branes}\label{sec:Localisedgravitons}

The metric of a magnetic brane resolved by transgression in $d$ dimensions that is sourced by a $(p+1)$-form potential has the form \cite{Cvetic:2000mh}
\begin{equation}\label{metricresolved}
ds^2_d = H^{-\frac{d_e}{d-2}}ds^2(\mb{R}^{1,d_m-1}) + H^{\frac{d_m}{d-2}}ds^2(B) \,,
\end{equation}
where $d_e = p+1$, $d_m = d-p-3$, and $B$ is a non-compact, $(p+3)$-dimensional, Ricci-flat manifold. The function $H$ satisfies
\begin{equation}
\Delta_BH = -\frac{1}{2k!}L^2_{(k)}\,,
\end{equation}
where $\Delta_B$ is the Laplacian on $B$, and $L_{(k)}$ is a harmonic $k$-form on $B$. Compared to the Salam-Sezgin solution of Type IIA supergravity, which can be interpreted as a resolved NS5-brane with $B$ being Eguchi-Hanson space, we find that the metric in \eqref{metricresolved} lacks a fibration of one of the worldvolume coordinates over $B$ (see \cite{Crampton:2014hia}). This does not affect our discussion here because we can always consider the same solution reduced along the fibre direction. For the Salam-Sezgin solution, this is equivalent to considering a reduction along the fibre coordinate to 9 dimensions, in which the resolved NS5-brane becomes a resolved 4-brane. Since the fibre direction is external to $B$, reducing on it does not introduce any curvature singularities.

What we are interested in is the behaviour of worldvolume gravitational waves on this background.  As demonstrated in \cite{Bachas:2011xa}, it is consistent to consider perturbations of the form
\begin{equation}
\eta_{\mu\nu} \mapsto \eta_{\mu\nu} + h_{\mu\nu}(x,y) \,,
\end{equation}
where $h_{\mu\nu}$ is transverse and traceless, and $y$ are the coordinates on $B$. The perturbation $h_{\mu\nu}$ satisfies the wave equation
\begin{equation}\label{beeqn}
\Box_dh_{\mu\nu} = 0 \,,
\end{equation}
where $\Box_d$ is the scalar Laplacian associated to the metric \eqref{metricresolved}. A little algebra shows that \eqref{beeqn} can be rewritten as
\begin{equation}
\left(\partial^2 + H^{-1}\Delta_B\right)h_{\mu\nu} = 0 \,,
\end{equation}
where $\partial^2 = \eta^{\mu\nu}\partial_\mu\partial_\nu$. The lower-dimensional physics is then controlled by the spectrum of the operator $H^{-1}\Delta_B$. In particular, we can expand $h_{\mu\nu}$ into eigenmodes of $H^{-1}\Delta_B$, 
\begin{equation}\label{expansion}
h_{\mu\nu}(x,y) = h^{(m)}_{\mu\nu}(x)\xi_{(m)}(y)\,,
\end{equation}
where 
\begin{equation}\label{eigen}
\Delta_B\xi_{(m)} = -m^2H\xi_{(m)} \,,
\end{equation}
and
\begin{equation}
(\partial^2-m^2)h^{(m)}_{\mu\nu} = 0\,.
\end{equation}
From this, we observe that in order for there to be an isolated, massless graviton on the worldvolume, the spectrum of $H^{-1}\Delta_B$ must contain a normalisable zero mode $\xi_{(0)}(y)$. To determine the precise measure with which to normalise $\xi_{(0)}(y)$, we can look at the $h^2$ terms in the perturbative action. This is most easily done by changing to a frame where the metric is a direct product,
\begin{equation}
ds^2_d = H^{-\frac{d_e}{d-2}}d\tilde s^2_d \,,\quad d\tilde s^2_d = (\eta_{\mu\nu}+ h_{\mu\nu})dx^\mu dx^\nu+ Hds^2(B) \,.
\end{equation}
The Einstein-Hilbert term under this change of frame is
\begin{equation}
\sqrt{-g}R = H^{-d_e/2}\left(\sqrt{-\tilde g}\tilde R + \cdots\right) \,.
\end{equation}
From this, we can read off the measure for integration over $B$ to be
\begin{equation}\label{measure}
\mu(y) = H^{-d_e/2}H^{(d_e+2)/2}\sqrt{g_B} = H\sqrt{g_B} \,,
\end{equation}
where $g_B$ is the metric on $B$. Although we have derived this measure by analysing the perturbative action, it is easy to see, from a differential equation point of view, that this is indeed the correct measure. The defining equation for the eigenmodes is \eqref{eigen}. Assuming appropriate boundary conditions on the eigenfunctions to ensure the self-adjointness of $\Delta_B$, \eqref{eigen} takes the form of a Sturm-Liouville equation on $B$ with weight function $H$. The Sturm-Liouville theorem then states that the measure with respect to which the eigenmodes are orthonormalised is the product of the weight function and the natural measure on $B$, which indeed agrees with \eqref{measure}. 

In the following section, we will review some mathematical theory concerning differential equations of the type \eqref{eigen}, and apply these results to several supergravity solutions, and will show how they can realise worldvolume gravity localisation. We will also consider an example of a supergravity solution where worldvolume gravity localisation cannot be realised.

\section{Some Sturm-Liouville theory}\label{sec:SLT}

Hermann Weyl began the analysis of Sturm-Liouville systems with singular endpoints in connection with the geometric construction of concentric circles in the plane, each one contained within the preceding one \cite{Weyl:1910}. He introduced the terminology of limit circles (LC) and limit points (LP) to describe the nature of the singularity at an endpoint. The analysis for
Sturm-Liouville problems is much more general, however and was developed in particular by A. Zettl and collaborators \cite{Zettl:2005}.

All of the examples that we will consider here consist of transverse geometries that have one distinguished non-compact direction. Let $z$ be this non-compact coordinate normalised such that $z\in[0,\infty)$. Since there is only one non-compact direction, we may integrate over the transverse-space compact directions so that the remaining  coordinate dependence of the eigenfunctions $\xi_{(m)}$ is only on the coordinate $z$. One may also view this as restricting to a standard consistent truncation for the compact transverse directions. Then the eigenvalue equation \eqref{eigen} becomes a Sturm-Liouville problem (SLP),
\begin{equation}\label{SLeigenreal}
\mc{L}\xi_{(m)} \coloneqq -(p\,\xi_{(m)}')' = m^2\mu\, \xi_{(m)} \,,\quad z \in[0,\infty) \,,
\end{equation}
where $p(z) > 0$, $\mu(z)>0$ is the measure, and the primes denote $z$ derivatives. The combination $p\,\xi_{(m)}'$ is called the quasi-derivative of $\xi_{(m)}$ with respect to the SLP \eqref{SLeigenreal}, or just the quasi-derivative for short. As opposed to the general canonical form of an SLP, there is no undifferentiated term on the LHS of \eqref{SLeigenreal} proportional to $\xi_{(m)}$. This is because the differential operator in \eqref{eigen} is the Laplacian, which has a $d(\cdot d)$ structure. For completeness, we also will extend the analysis to the case where $m$ can be imaginary in the following section, so that the SLP becomes
\begin{equation}\label{SLeigen}
\mc{L}\xi_{(\lambda)} \coloneqq -(p\,\xi_{(\lambda)}')' = \lambda \mu\, \xi_{(\lambda)} \,,\quad \lambda\in\mb{R}\,,\quad z \in[0,\infty) \,.
\end{equation}
In the following examples, the endpoint $z=\infty$ will be singular while $z=0$ will either be regular or singular. This makes the problem of defining boundary conditions somewhat intricate because the solutions to \eqref{SLeigen} and their quasi-derivatives are, in general, undefined at the singular points. A consequence of this is that a particular self-adjoint realisation $\mc{L}^s$ of the SL operator $\mc{L}$ needs to be constructed. Only when $\mc{L}$ is made self-adjoint by $\mc{L}^s$ does one enjoy desirable spectral properties such as the reality of eigenvalues and the orthogonality of eigenfunctions with different eigenvalues. 

There are, in general, many self-adjoint realisations of $\mc{L}$. The differences between them boil down to the boundary conditions imposed on the endpoints, which are in turn dependent on the nature of the endpoints. We then need to understand the different types of endpoint as well as how they influence the construction of self-adjoint realisations of $\mc{L}$. This will be the focus of this subsection, following the presentation given in Ref. \cite{Zettl:2005}. 

Before diving into the different types of singular endpoints, we note that the self-adjoint domain of $\mc{L}^s$, for any type of singular endpoint, is a subset of the maximal domain 
\begin{equation}
D_{\text{max}} \coloneqq \{\xi: [0,\infty)\to\mb{R}: \xi, p\xi' \in\Omega^0((0,\infty))\,, \xi, \mu^{-1}\mc{L}\xi \in L^2([0,\infty),\mu)\} \,,
\end{equation}
which is characterised by boundary conditions involving linear combinations of the Lagrange sesquilinear form 
\begin{equation}
[f,g] \coloneqq  f(pg') - g(pf') 
\end{equation}
defined at one or both endpoints. The space $L^2([0,\infty),\mu)$ of square-integrable functions with respect to the measure $\mu$ is the Hilbert space $\mc{H}$,
\begin{equation}
\mc{H} =  L^2([0,\infty),\mu) \,.
\end{equation}
We note that the final condition in the definition of the maximal domain, $\mu^{-1}\mc{L}\xi \in \mc{H}$, is implied by the penultimate condition, $\xi\in \mc{H}$, for all $\xi$ solving \eqref{SLeigen}. 

There are many classifications of singular endpoints. Classifying whether a singular endpoint is a regular singular point or an irregular singular point, for example, is useful for performing a Frobenius analysis. For our purposes, however, it is more useful to classify the singular endpoints as either limit-circle (LC) or limit-point (LP). An endpoint is LC if all solutions to \eqref{SLeigen} are square-integrable with respect to the measure $\mu$ in a neighbourhood of the endpoint. If a singular endpoint is not LC, it is LP. From this definition, the LC/LP classification of an endpoint appears to depend on the eigenvalue $\lambda$. This, however, is not the case due to the following theorem:\footnote{See Theorem 7.2.2 in Ref.\ \cite{Zettl:2005}.} 
\begin{theorem}
If all solutions of \eqref{SLeigen} for a particular $\lambda\in\mb{R}$ are in $L^2(I,\mu)$ for some interval $I\subset[0,\infty)$, then this holds for every $\lambda$. 
\end{theorem}
Consequently, we can classify whether an endpoint is LC or LP by looking at the solutions to \eqref{SLeigen} for a particular value of $\lambda$. The simplest solutions are the zero modes, corresponding to $\lambda=0$. Since \eqref{SLeigen} does not have a term proportional to $\xi_{(\lambda)}$ on the LHS, the zero modes can be obtained by directly integrating the quasi-derivative,  
\begin{equation}
\xi_{(0)}(z) = a_{(0)} + b_{(0)}\int^z\frac{dt}{p(t)} \coloneqq \xi^{c}_{(0)} + \xi^{nc}_{(0)}(z) \,,
\end{equation}
where $a_{(0)}$ and $b_{(0)}$ are constants, and the superscripts $c$ and $nc$ stand for constant and non-constant. The condition for the endpoint $z=0$ to be LC is then
\begin{equation}
\int_0^b dz\,\mu(z) < \infty \,,\quad \text{and}\quad \int_0^b dz\,\mu(z)\xi^{nc}_{(0)}(z)^2 < \infty \,,
\end{equation}
where $b$ is a non-zero constant. Similar conditions are made at $z=\infty$. For the examples below that admit worldvolume gravity localisation, the point $z=0$ is either regular or LC, while the point $z=\infty$ is LP. In such cases\footnote{See Theorem 10.4.5 in Zettl \cite{Zettl:2005}.}, for any pair of non-zero real numbers $(A_1,A_2)$,
\begin{equation}\label{selfadjointdomain}
D^s(\omega, A_1,A_2) \coloneqq \{\xi \in D_{\text{max}} : A_1[\xi,u_{(\omega)}](0) + A_2[\xi,v_{(\omega)}](0) = 0 \} 
\end{equation}
is a self-adjoint domain, where $u_{(\omega)}$ and $v_{(\omega)}$ are the linearly independent solutions to \eqref{SLeigen} with $\lambda=\omega$. For $\omega=0$, these are $\xi^c_{(0)}$ and $\xi^{nc}_{(0)}$. In other words, the self-adjoint extension $\mc{L}^s$ of $\mc{L}$ with domain $D$ is constructed by imposing the boundary condition
\begin{equation}\label{selfadjointbc}
A_1[\xi_{(\lambda)},u_{(\omega)}](0) + A_2[\xi_{(\lambda)},v_{(\omega)}](0) = 0 \,,\quad (A_1, A_2) \in\mb{R}^2\setminus\{(0,0)\} \,.
\end{equation}
Importantly, there are no special conditions other than $L^2$ normalisation required at the LP point, $z=\infty$. All of the required information is encoded in the above boundary condition at the LC point $z=0$. \\

With the self-adjoint realisation $\mc{L}^s$ of $\mc{L}$ constructed with the domain \eqref{selfadjointdomain}, we can now study its spectrum. Since $\mc{L}^s$ is self-adjoint, the spectrum $\sigma(\mc{L}^s)$ is a subset of $\mb{R}$, and is the union of two disjoint sets
\begin{equation}
\sigma(\mc{L}^s) = \sigma_d(\mc{L}^s) \cup \sigma_{\text{ess}}(\mc{L}^s) \,,
\end{equation}
where $\sigma_d(\mc{L}^s) $ is the discrete spectrum and $\sigma_{\text{ess}}(\mc{L}^s)$ is the essential spectrum. The discrete spectrum consists of all isolated eigenvalues of finite multiplicity. and their associated eigenfunctions are square-integrable. In particular, $\xi_{(0)}$ is square-integrable. On the other hand, the essential spectrum consists of real numbers $\lambda$ such that the solutions to \eqref{SLeigen} are not square-integrable. In the language of quantum mechanics, the discrete spectrum corresponds to bound states, while the essential spectrum corresponds to a continuum of scattering states. 

In the cases that we will be considering, where $z=0$ is either regular or LC and $z=\infty$ is LP, the essential spectrum does not depend on the particular boundary condition \eqref{selfadjointbc}, or in other words, on the particular self-adjoint realisation of $\mc{L}$, but the discrete spectrum does.\footnote{See Proposition 10.4.4 (2) in Zettl \cite{Zettl:2005}.} We will thus abbreviate the notation for the essential spectrum to $\sigma_{\text{ess}}$. Let 
\begin{equation}
\sigma_0 \coloneqq \inf\sigma_{\text{ess}} 
\end{equation}
be the starting point of the essential spectrum. The possible spectra depend on the following three possibilities of $\sigma_0$: \\

\noindent \textbf{1.} $\sigma_0=-\infty$. In this case, $\sigma_{\text{ess}}$ may be the whole real line, or it may consist of disjoint closed intervals separated by gaps. Given an eigenvalue $\lambda\in \sigma_{\text{ess}}$, its corresponding eigenfunction has an infinite number of zeroes in $[0,\infty)$. A function with an infinite number of zeroes is called oscillatory, whereas a function with a finite number of zeroes is called non-oscillatory. \\

\noindent \textbf{2.} $\sigma_0=\infty$. This is the case where $\sigma_{\text{ess}}$ is empty, so the spectrum of every self-adjoint realisation is discrete, $\sigma(\mc{L}^s) = \sigma_d(\mc{L}^s)$, and there are only bound states. \\

\noindent \textbf{3.} $\sigma_0\in(-\infty,\infty)$. Here, the point $\sigma_0$ is called the oscillation number. For $\lambda<\sigma_0$, each nontrivial solution has no zeroes or a finite number of zeroes. If $\lambda>\sigma_0$, then each nontrivial solution has an infinite number of zeroes. If $\lambda=\sigma_0$, then either of the two cases above can occur. \\

In the following, we will employ the Sturm-Liouville analysis reviewed here to two known examples (Randall-Sundrum and Salam-Sezgin), and a few new examples of brane gravity localisation. For all of these examples, the self-adjoint domain will be chosen with respect to the zero eigenfunction. Recall that a bound-state solution to \eqref{SLeigen} is an eigenfunction if it is in the Hilbert space $\mc{H}$, \ie if it is square-integrable. Since $z$ is a non-compact coordinate, the zero eigenfunction is not necessarily constant. In any case, the boundary condition at $z=0$ is 
\begin{equation}\label{zeroselfadjointbc}
[\xi_{(\lambda)},\xi_{(0)}](0) = 0 \,,
\end{equation}
where $\xi_{(0)}$ is the zero eigenfunction. We will denote the self-adjoint realisation of $\mc{L}$ with the domain defined by \eqref{zeroselfadjointbc} by $\mc{L}^0$. Note that $\xi_{(0)}$ itself automatically obeys this boundary condition. The solutions to \eqref{SLeigen} obeying the boundary condition \eqref{zeroselfadjointbc} are orthogonal to $\xi_{(0)}$. In particular, the square-integrable bound states will satisfy 
\begin{equation}\label{boundortho}
\int_0^\infty dz\,\mu\,\xi_{(\lambda)}\xi_{(\omega)} = \delta_{\lambda\omega} \,,\quad \lambda,\omega\in\sigma_d(\mc{L}^0) \,,
\end{equation}
while the scattering states obey
\begin{equation}\label{scatteringortho}
\int_0^\infty  dz\,\mu\,\xi_{(\lambda')}\xi_{(\omega')} = \delta(\lambda'-\omega') \,,\quad \lambda',\omega'\in\sigma_{\text{ess}} \,.
\end{equation}
Consequently, we can use \eqref{boundortho} as an easy way to see which eigenvalues are allowed in $\sigma_d(\mc{L}^0)$. Any solution to \eqref{SLeigen} with $\lambda\notin\sigma_{\text{ess}}$ that is not square-integrable is thus not part of the spectrum. In some cases, it may be difficult to compute the integral \eqref{boundortho} explicitly. However, for $\omega\neq\lambda$ and $\omega\neq0$, a little algebra shows that 
\begin{equation}
0 = \int_0^\infty dz\,\mu\,\xi_{(\lambda)}\xi_{(\omega)} = \frac{\lambda}{\omega}\int_0^\infty dz\,\mu\,\xi_{(\lambda)}\xi_{(\omega)} - \frac{1}{\omega}[\xi_{(\lambda)},\xi_{(\omega)}]\Big|_{0}^{\infty} \,,
\end{equation}
which is equivalent to
\begin{equation}
[\xi_{(\lambda)},\xi_{(\omega)}]\Big|_{0}^{\infty} = 0 \,.
\end{equation}
Since $\xi_{(\lambda)}$ and $\xi_{(\omega)}$ obey \eqref{zeroselfadjointbc}, the condition at $z=0$ is satisfied trivially. Therefore, the condition for orthonormality is then
\begin{equation}
[\xi_{(\lambda)},\xi_{(\omega)}](\infty) = 0 \,.
\end{equation}
for all pairs $\xi_{(\lambda)}$ and $\xi_{(\omega)}$, $\omega\neq\lambda$. This shows that only the asymptotic behaviour of the solutions at infinity is required to check whether they belong in the spectrum. 

In the following, the supergravity solution for Example 1 is from \cite{Randall:1999vf}, Example 2 is from \cite{Crampton:2014hia}, Examples 3-5 are from \cite{Vazquez-Poritz:2012yhv}, and Example 6 is from \cite{Cvetic:2000mh}.

\section{Examples}\label{sec:examples}
\subsection{Example 1: Randall-Sundrum}

Although the Randall-Sundrum (RSII) geometry is not that of a resolved brane, but rather that of an orbifolded $\hbox{AdS}_5$ sourced by a codimension 1 brane, we can still understand the brane gravity localisation in this case using the above framework. The relevant SLP is\footnote{We have shifted the $z$ coordinate of \cite{Randall:1999vf} by $1/k$ so that it starts at 0.} 
\begin{equation}\label{RSSLP}
\xi_{(\lambda)}'' - \frac{3}{(z+1/k)}\xi_{(\lambda)} = -\lambda \xi_{(\lambda)} \,,
\end{equation}
where $k>0$ is a constant, $\lambda\in\mb{R}$, and $z\in[0,\infty)$. The case of physical interest is when $\lambda=m^2>0$, but we will first perform an analysis for all real $\lambda$. Equation \eqref{RSSLP} can be written in the form of \eqref{SLeigen} with $p(z) = (z+1/k)^{-3}$ and $\mu(z) = p(z)$. The endpoint $z=0$ is regular, but $z=\infty$ is singular, and is LP. To see that it is LP, consider the zero mode solutions to \eqref{RSSLP},
\begin{equation}\label{RSzero}
\xi_{(0)}(z) = a_{(0)} + b_{(0)}(z+1/k)^4 = \xi^c_{(0)} +  \xi^{nc}_{(0)}(z)\,.
\end{equation}
We have $\mu\, \xi^c_{(0)} \propto (z+1/k)^{-3}$ and $\mu\,\xi^{nc}_{(0)}\propto (z+1/k)$. Thus, $\xi^c_{(0)}$ is square-integrable in the neighbourhood of $z=\infty$, but $\xi^{nc}_{(0)}$ is not. Hence, $z=\infty$ is LP. This also shows that the zero eigenfunction is the constant mode
\begin{equation}
\xi_{(0)}(z) = a_{(0)} \,.
\end{equation}
The general solution to \eqref{RSSLP} is given by
\begin{equation}
\xi_{(\lambda)}(z) = (z+1/k)^2\left(a_{(\lambda)}J_2\left(\sqrt{\lambda}(z+1/k)\right) + b_{(\lambda)}Y_2\left(\sqrt{\lambda}(z+1/k)\right)\right)\,,
\end{equation}
where $a_{(\lambda)}$ and $b_{(\lambda)}$ are constants, and $J$ and $Y$ are Bessel functions of the first and second kind respectively. These functions are oscillatory (scattering) for all $\lambda>0$, so the essential spectrum is $\sigma_{\text{ess}} = (0,\infty)$, and the oscillation number is 
\begin{equation}
\sigma_0 = \inf\sigma_{\text{ess}} = 0 \,.
\end{equation}
To find the discrete spectrum, we need to impose the boundary condition \eqref{zeroselfadjointbc} at $z=0$. Since the zero mode is constant, this is equivalent to the Neumann condition
\begin{equation}\label{RSbc}
\xi_{(\lambda)}'(0) = 0 \,.
\end{equation}
For $\lambda = m^2>0$, we have
\begin{equation}
b_{(\lambda)} = -\frac{J_1(\sqrt{\lambda}/{k})}{Y_1(\sqrt{\lambda}/{k})}a_{(\lambda)} = -\frac{J_1(m/{k})}{Y_1(m/{k})}a_{(\lambda)} \,.
\end{equation}
For $\lambda<0$, the solutions obeying \eqref{RSbc} are not square-integrable, so they are not part of the discrete spectrum. Thus, we find that the orthonormal solutions to \eqref{RSSLP} in the self-adjoint domain defined by \eqref{RSbc} are given by
\begin{equation}
\xi_{(0)} = a_{(0)} \,,\quad \xi_{(m)} = a_{(m)}(z+1/k)^2\left(J_2\left(m(z+1/k)\right) -\frac{J_1(m/{k})}{Y_1(m/{k})}Y_2\left(m(z+1/k)\right)\right) \,,
\end{equation}
with $\lambda=m^2$. The coefficients $a_{(0)}$ and $a_{(m)}$ are determined by the normalisations
\begin{equation}
\int_0^\infty dz\,\mu\,\xi^2_{(0)} = 1 \,,\quad \int_0^\infty dz\,\mu\,\xi_{(m)}\xi_{(n)} = \delta(\lambda(m)-\lambda(n)) = \delta(m^2-n^2) = \frac{\delta(m-n)}{2n} \,.
\end{equation}
For the zero eigenfunction, we have
\begin{equation}
1 = \int_0^\infty dz\,\mu\,\xi^2_{(0)} = \frac{a_{(0)}^2k^2}{2} \,.
\end{equation}

\subsection{Example 2: Salam-Sezgin}

The Salam-Sezgin $\mb{R}^{1,3}\times S^2$ vacuum lifted to Type IIA supergravity has the form of a completely resolved NS5-brane on Eguchi-Hanson space. The metric on the Eguchi-Hanson space is given by \eqref{EHmetric}. By relabelling $z = \rho$, the relevant SLP is
\begin{equation}\label{SSSLP}
\xi''_{(\lambda)} + 2\coth2z\,\xi'_{(\lambda)} = -\lambda\,\xi_{(\lambda)} \,,
\end{equation}
which can be put into the form \eqref{SLeigen} with $p(z)=\mu(z)=\sinh2z$. The endpoint $z=0$ is LC, whereas $z=\infty$ is LP. To see this, note that the zero modes are given by
\begin{equation}
\xi_{(0)}(z) = a_{(0)} + b_{(0)}\log\tanh z = \xi^c_{(0)} + \xi^{nc}_{(0)}(z) \,.
\end{equation}
These are both square-integrable with respect to the measure $\mu=\sinh2z$ in a neighbourhood of $z=0$, but $\xi^{nc}_{(0)}$ is singular at $z=0$, so $z=0$ is LC. The constant mode $\xi^c_{(0)}$ is clearly not square-integrable in a neighbourhood of $z=\infty$, so $z=\infty$ is LP. To deduce the essential spectrum, we consider the asymptotic form of \eqref{SSSLP} near $z=\infty$. This is
\begin{equation}
\xi''_{(\lambda)} + 2\xi'_{(\lambda)} = -\lambda\,\xi_{(\lambda)} \,,
\end{equation}
which has solutions 
\begin{equation}
\xi_{(\lambda)} = e^{-\rho}\left(a_{(\lambda)}\cos(\sqrt{\lambda-1}z) +b_{(\lambda)}\sin(\sqrt{\lambda-1}z)\right)\,.
\end{equation}
These are oscillatory for $\lambda > 1$, so the essential spectrum is $\sigma_{\text{ess}} = (1,\infty)$, and the oscillation number is
\begin{equation}
\sigma_0 = \inf\sigma_{\text{ess}} = 1 \,.
\end{equation}
To deduce the discrete spectrum, we impose the boundary condition \eqref{zeroselfadjointbc} at $z=0$, 
\begin{equation}
[\xi_{(\lambda)},\xi_{(0)}](0) = 0 \,,
\end{equation}
where the square-integrable zero eigenfunction $\xi_{(0)}$ is given by
\begin{equation}
\xi_{(0)}(z) = b_{(0)}\log\tanh z \,.
\end{equation}
Explicitly, the boundary condition is
\begin{equation}\label{SSbc}
\lim_{z\to0^+}\left(\sinh2z\,\log\tanh z\, \xi'_{(\lambda)}(z) - 2\xi_{(\lambda)}(z)\right) = 0 \,,
\end{equation}
where we used the fact that $\sinh2z\,\xi'_{(0)}(z) = 2b_{(0)}$. To see which bound states other than the zero eigenfunction are allowed, we note that the general solution to \eqref{SSSLP} for arbitrary $\lambda$ is
\begin{equation}
\xi_{(\lambda)}(z) = a_{(\lambda)}\mc{Q}_{-\frac{1}{2}-\frac{\sqrt{1-\lambda}}{2}}(\cosh2z) + b_{(\lambda)}\mc{Q}_{-\frac{1}{2}+\frac{\sqrt{1-\lambda}}{2}}(\cosh2z) \,,
\end{equation}
where $\mc{Q}$ is the Legendre function of the second kind. Since $\lambda < 1$ for the (possible) bound states, the orders of the Legendre functions are real, and we define $\nu_1 = -\tfrac{1}{2}-\tfrac{\sqrt{1-\lambda}}{2}$ and $\nu_2 = -\tfrac{1}{2}+\tfrac{\sqrt{1-\lambda}}{2}$. The boundary condition at $z=0$ then sets
\begin{equation}
b_{(\lambda)} = -\frac{H(\nu_1)}{H(\nu_2)}a_{(\lambda)} \,,
\end{equation}
where $H(n) \coloneqq \int_0^1 dt\,(1-t^n)/(1-t)$ is Euler's representation of the harmonic number $H(n)$.\footnote{This relation is also true for the scattering states.} Although the boundary condition at $z=0$ can be satisfied, solutions with $\lambda<1$ and $\lambda\neq0$ are not normalisable. Thus, they are not in the Hilbert space $\mc{H}$, and are therefore not part of the spectrum. So, we find that the orthonormal solutions to \eqref{SSSLP} in the self-adjoint domain defined by \eqref{SSbc} are given by
\begin{equation}
\begin{split}
&\xi_{(0)} = b_{(0)}\log\tanh z \,,\quad \\
&\xi_{(m)} = a_{(m)}\left(\mc{Q}_{-\frac{1}{2}-\frac{i\sqrt{m^2-1}}{2}}(\cosh2z) -\frac{H(\nu_1)}{H(\nu_2)}\mc{Q}_{-\frac{1}{2}+\frac{i\sqrt{m^2-1}}{2}}(\cosh2z)\right) \,,\quad m>1\,,
\end{split}
\end{equation}
with $\lambda=m^2$, $\nu_1 =  -\tfrac{1}{2}-\tfrac{i\sqrt{m^2-1}}{2}$, and $\nu_2 = \nu_1^*$. The coefficients $b_{(0)}$ and $a_{(m)}$ are determined by the normalisations
\begin{equation}
\int_0^\infty dz\,\mu\,\xi^2_{(0)} = 1 \,,\quad \int_0^\infty dz\,\mu\,\xi_{(m)}\xi_{(n)} =  \frac{\delta(m-n)}{2n} \,.
\end{equation}
For the zero eigenfunction, we have
\begin{equation}
1 = \int_0^\infty dz\,\mu\,\xi^2_{(0)} = \frac{b_{(0)}^2\pi^2}{12} \,.
\end{equation}

\subsection{Example 3: D3-branes on a resolved conifold over $S^5/\mb{Z}_3$}

The transverse metric is given by 
\begin{equation}\label{D31transverse}
\begin{split}
&ds^2(B) = f^{-1}dr^2 + \frac{r^2}{9}f\left(d\psi - 3\sin^2\theta \sigma_3\right)^2 + r^2 ds^2(\mb{C}\text{P}^2) \,,\quad f = 1- \frac{b^6}{r^6} \,,\\
&ds^2(\mb{C}\text{P}^2) = d\theta^2 + \sin^2\theta\left(\sigma^2_1 + \sigma_2^2 + \cos^2\theta\,\sigma_3^2\right) \,,
\end{split}
\end{equation}
where $r\geq b$, and $\sigma_i$ are the left-invariant 1-forms of $SU(2)$ satisfying
\begin{equation}
d\sigma_i = 2\epsilon_{ijk}\sigma_j\wedge\sigma_k  \,.
\end{equation}
The corresponding function $H$ is given by
\begin{equation}\label{Hex3}
H = \frac{1}{4b^6r^4} \,.
\end{equation}
To understand the geometry as $r\to b^+$, consider the coordinate transformation 
\begin{equation}
r = b(\cosh\rho)^{1/3} \,.
\end{equation}
Then, the metric is
\begin{equation}
ds^2(B) = \frac{b^2}{9}(\cosh\rho)^{2/3}\left(d\rho^2 + (\tanh\rho)^2\left(d\psi - 3\sin^2\theta \sigma_3\right)^2 + 9ds^2(\mb{C}\text{P}^2)\right) \,.
\end{equation}
Near $\rho = 0$, this is
\begin{equation}
ds^2(B)_{\rho\to0^+} = \frac{b^2}{9}\left(d\rho^2 + \rho^2\left(d\psi - 3\sin^2\theta \sigma_3\right)^2 + 9ds^2(\mb{C}\text{P}^2)\right) \,,
\end{equation}
so $B$ is asymptotically $\mb{R}^2\times \mb{C}\text{P}^2$ provided $\psi\in[0,2\pi)$. As such, we expect the zero mode to exhibit a logarithmic behaviour near the origin. 

The variables $r$ and $\rho$ are useful in understanding the geometry of the transverse space. However, to analyse the eigenvalue equation \eqref{eigen}, it is more useful to consider the following coordinate redefinition
\begin{equation}
r = b (z+1)^{1/2} \,,
\end{equation}
where $z \in[0,\infty)$. The SLP then reads
\begin{equation}\label{D31SLP}
z(z^2+3z+3)\xi''_{(\lambda)} + 3(z+1)^2\xi'_{(\lambda)} = -\frac{\lambda}{16b^8}\xi_{(\lambda)} \,.
\end{equation}
This can be put into the canonical form \eqref{SLeigen} with $p(z) = z(z^2+3z+3)$ and $\mu(z) = 1$. For the rest of this section, we will absorb the factor of $16b^8$ into $\lambda$. To classify whether the endpoints $z=0$ and $z=\infty$ are LC or LP, we note that the zero modes are
\begin{equation}\label{D31zero}
\xi_{(0)}(z) = a_{(0)} + b_{(0)}\left(2\sqrt{3}\tan^{-1}\left(\frac{\sqrt{3}}{3+2z}\right) + \log\left(\frac{z^2}{z^2+3z+3}\right)\right) = \xi^c_{(0)} + \xi^{nc}_{(0)}(z) \,.
\end{equation}
Both $\xi^c_{(0)}$ and $\xi^{nc}_{(0)}$ are square-integrable near $z=0$, but $\xi^{nc}_{(0)}$ is logarithmically divergent at this point, so $z=0$ is LC. Near $z=\infty$, $\xi^c_{(0)}$ is not square-integrable, so $z=\infty$ is LP. To determine the essential spectrum, consider the variable redefinition $z=1/y$. In the $y$ coordinate, the eigenvalue equation becomes
\begin{equation}
y(1+3y(1+y))\xi''_{(\lambda)} + (3y^2-1)\xi'_{(\lambda)} = -\lambda\xi_{(\lambda)} \,.
\end{equation}
In a finite interval near $y=0$, this becomes
\begin{equation}
\xi''_{(\lambda)} -\frac{1}{y}\xi'_{(\lambda)} = -\frac{\lambda}{y}\xi_{(\lambda)} \,,
\end{equation}
which has solutions 
\begin{equation}
\xi_{(\lambda)} = y\left(a_{(\lambda)} J_2\left(2\sqrt{\lambda y}\right) + b_{(\lambda)}Y_2\left(2\sqrt{\lambda y}\right)\right)\,.
\end{equation}
As in the Randall-Sundrum case, the Bessel functions $J_2$ and $Y_2$ are oscillatory for all $\lambda>0$. This means that the oscillation number is
\begin{equation}
\sigma_0 = \inf\sigma_{\text{ess}} = 0 \,,
\end{equation}
and there is no mass gap. 

We also note that in order for the zero mode \eqref{D31zero} to be square-integrable, we require $a_{(0)} = 0$. The zero eigenfunction, written back in the $z$ coordinate, is then
\begin{equation}
\xi_{(0)} = b_{(0)}\left(2\sqrt{3}\tan^{-1}\left(\frac{\sqrt{3}}{3+2z}\right) + \log\left(\frac{z^2}{z^2+3z+3}\right)\right) \,,
\end{equation}
with
\begin{equation}
\begin{split}
\int_0^\infty dz\,\xi^2_{(0)} &= \frac{b_{(0)}^2}{3}\left(\psi^{(1)}\left(\frac{1}{6}\right) + \psi^{(1)}\left(\frac{1}{3}\right) - \psi^{(1)}\left(\frac{2}{3}\right) - \psi^{(1)}\left(\frac{5}{6}\right)-2\pi^2\right) \\
&\approx 7.5\times b_{(0)}^2 \,,
\end{split}
\end{equation}
where $\psi^{(1)}(x)$ is the first derivative of the digamma function, $\psi(x) = \Gamma'(x)/\Gamma(x)$, and we recall that the measure $\mu(z)=1$.

The function $H$ given in \eqref{Hex3} is analogous to the one in \eqref{sourcedharmonicEH} with $k=0$. It is the function describing a fully resolved background with no singularity. We can similarly add to $H$ a term corresponding to the homogeneous solution to the Poisson equation that it satisfies, \ie a term proportional to $\xi_{(0)}$. This addition corresponds to an inclusion of a brane. Since $\xi_{(0)}$ is negative definite for all $z$, we write
\begin{equation}
H = \frac{1}{4b^{10}(z+1)^2} - k\left(2\sqrt{3}\tan^{-1}\left(\frac{\sqrt{3}}{3+2z}\right) + \log\left(\frac{z^2}{z^2+3z+3}\right)\right) \,,\quad k \geq 0 \,.
\end{equation}
With this new $H$, the endpoints $z=0$ and $z=\infty$ of the SLP are still LC and LP respectively, and near $y=1/z = 0$, the SLP becomes
\begin{equation}
\xi''_{(\lambda)} -\frac{1}{y}\xi'_{(\lambda)} = -\frac{\lambda(1+12b^{10}k)}{y}\xi_{(\lambda)} \,.
\end{equation}
The solutions to this SLP are scattering for all $k \geq 0$, so even with the inclusion of the brane, this system does not have a mass gap.

\subsection{Example 4: D3-branes on resolved conifolds over $Y^{p,q}$}

The metric on the transverse space is a resolved cone over the Sasaki-Einstein spaces $Y^{p,q}$ constructed in \cite{Gauntlett:2004yd},
\begin{equation}\label{D32transverse}
ds^2(B) = f^{-1}dr^2+\frac{1}{9}fr^2(d\psi+\mc{A}_{(1)})^2 + r^2ds^2_4 \,,
\end{equation}
where
\begin{equation}
f = 1 - \frac{b^6}{r^6} \,,\quad \mc{A}_{(1)} = -\cos\theta\,d\varphi + y(d\beta + \cos\theta\,d\varphi) 
\end{equation}
and the K\"{a}hler-Einstein base is 
\begin{equation}
ds^2_4 = \frac{1}{6}(1-y)(d\theta^2 + \sin^2\theta\,d\varphi^2) + \frac{dy^2}{w(y)q(y)}+ \frac{1}{36}w(y)q(y)(d\beta + \cos\theta\,d\varphi)^2\,,
\end{equation}
with
\begin{equation}
w(y) = \frac{2(a-y^2)}{1-y} \,,\quad q(y) = \frac{a-2y^2+2y^3}{a-y^2} \,.
\end{equation}
The coordinate ranges are $r\geq b$, $\theta\in[0,\pi]$, $\varphi\in[0,2\pi)$, and $y\in[y_1,y_2]$ with
\begin{equation}
a = \frac{1}{2} - \frac{p^2-3q^2}{4p^3}\sqrt{4p^2-3q^2}\,,\quad y_1 = \frac{1}{4p}\left(2p-3q- \sqrt{4p^2-3q^2}\right)\,,\quad y_2 =y_1 + \frac{3q}{2p} \,.
\end{equation}
The short-distance geometry is $\mb{R}^2\times KE_4$, where $KE_4$ is the K\"{a}hler-Einstein base, and the requisite function $H$ is the same as for the previous example,
\begin{equation}
H = \frac{1}{4b^6r^4} \,.
\end{equation}
For functions that only depend on $r$, the Laplacian on the space \eqref{D32transverse} is the same as that on \eqref{D31transverse}. Thus, the relevant eigenvalue problem is just \eqref{D31SLP} again, and the same analysis from the previous section goes through.

\subsection{Example 5: D3-branes on a resolved cone over $T^{1,1}/\mb{Z}_2$}

The transverse metric is
\begin{equation}
ds^2(B) = f^{-1}dr^2 + \frac{r^2}{36}f(\sigma_3-\Sigma_3)^2 + \frac{1}{12}R_1^2(\sigma_1^2+\sigma_2^2) + \frac{1}{12}R_2^2(\Sigma_1^2+\Sigma_2^2) \,,
\end{equation}
where $\sigma_i$ and $\Sigma_i$ are left-invariant 1-forms of $SU(2)\times SU(2)$, $R^2_i = r^2+l_i^2$, $r\geq0$ and
\begin{equation}
f = \frac{2r^4+3(l_1^2+l_2^2)r^2+6l_1^2l_2^2}{R_1^2R_2^2} \,.
\end{equation}
Near $r=0$, the geometry becomes $\mb{R}^2\times S^2\times S^2$. Let's focus on the case $l_1 = l_2 = l$ with $R_1 = R_2 = R$, and with the normalisable, harmonic 2-form
\begin{equation}\label{D33harmonic2}
L_{(2)} = u_0 \theta^0 \wedge \theta^5 + u_1 \theta^1\wedge\theta^2 + u_2\theta^3\wedge\theta^4\,,
\end{equation}
where
\begin{equation}
u_0 = \frac{2}{R^6}\,,\quad u_1 = \frac{1}{R^6}\,,\quad u_2 = -\frac{1}{R^6} \,,
\end{equation}
and where the vielbeins are
\begin{equation}
\begin{split}
&\theta^0 = f^{-1/2}dz\,,\quad \theta^1 = \frac{R}{\sqrt{12}}\sigma_1 \,,\quad \theta^2 = \frac{R}{\sqrt{12}}\sigma_2\,,\\
&\theta^3 = \frac{R}{\sqrt{12}}\Sigma_1 \,,\quad \theta^4 = \frac{R}{\sqrt{12}}\Sigma_2 \,,\quad \theta^5 = \frac{r}{6}f^{1/2}(\sigma_3-\Sigma_3) \,.
\end{split}
\end{equation}
The requisite function $H$ that is sourced by \eqref{D33harmonic2} is given by
\begin{equation}
H = \frac{1}{8l^6R^4} \,,
\end{equation}
and the relevant eigenvalue equation is
\begin{equation}\label{D33SLP}
\xi''_{(\lambda)} + \frac{5r^4+9l^2r^2+3l^4}{r(r^4+3l^2r^2+3l^4)}\xi'_{(\lambda)} = -\frac{\lambda}{16l^6}\frac{1}{r^4+3l^2r^2+3l^4}\xi_{(\lambda)} \,.
\end{equation}
Absorbing the factor of $16l^6$ into $\lambda$, this can be put into the canonical form of \eqref{SLeigen} with $p(r) = r(r^4+3l^2r^2+3l^4)$, and $\mu(r) = r$. In fact, by performing the coordinate redefinition
\begin{equation}
r=l\sqrt{z} \,,\quad z\in[0,\infty) \,,
\end{equation}
we find that \eqref{D33SLP} becomes again \eqref{D31SLP}. Thus, the same analysis for \eqref{D31SLP} holds in this case as well.

\subsection{Example 6 (no localisation): NS5-branes on Taub-NUT}

The metric on the Taub-NUT space is given by 
\begin{equation}
ds^2(B) = \left(\frac{r+b}{r-b}\right)dr^2 + 4b^2 \left(\frac{r-b}{r+b}\right)(d\psi+\cos\theta\,d\varphi)^2 + (r^2-b^2)(d\theta^2+\sin^2\theta\,d\varphi^2) \,,
\end{equation}
where $r\geq b$, $\psi \in [0,4\pi)$, $\theta\in[0,\pi]$ and $\varphi\in[0,2\pi)$. Near $r=b$, the space is asymptotically $\mb{R}^4$, whereas near $r=\infty$, it becomes $\mb{R}^3\times S^1$. From \cite{Cvetic:2000mh}, the requisite function $H$ is 
\begin{equation}
H = \frac{1}{4b(r+b)} \,.
\end{equation}
Defining $z = r-b \in [0,\infty)$, the relevant SLP is then
\begin{equation}
\xi''_{(\lambda)} + \frac{2}{z}\xi'_{(\lambda)} = -\frac{\lambda}{4b z}\xi_{(\lambda)} \,,
\end{equation}
which can be put into the form of \eqref{SLeigen} with $p(z) = z^2$ and $\mu(z) = z$. The zero modes are given by\footnote{The radial coordinate of the $\mb{R}^4$ asymptote of the Taub-NUT space is $u=\sqrt{z}$, hence the $1/z$ structure of the zero modes.}
\begin{equation}
\xi_{(0)}(z) = a_{(0)} + \frac{b_{(0)}}{z} = \xi^c_{(0)} + \xi^{nc}_{(0)}(z) \,.
\end{equation}
$\xi^c_{(0)}$ is square-integrable near $z=0$, but $\xi^{nc}_{(0)}$ is not, so $z=0$ is LP. Similarly, $\xi^{nc}_{(0)}$ is square-integrable near $z=\infty$, but $\xi^{c}_{(0)}$ is not, so $z=\infty$ is also LP. Since both endpoints are LP, it is clear that there exists no zero eigenfunction. As such, a resolved NS5-brane on the Taub-NUT space does not admit worldvolume gravity localisation.

%

\vspace{-1em}
\section{Conclusion and a Conjecture}\label{sec:Concl}

The analysis of Sturm-Liouville problems with singular endpoints in the LC-LP language of limit circles and limit points clearly identifies the asymptotic properties of brane transverse wave equations that allow gravity localisation in the region near a brane worldvolume. As explored in Ref.\ \cite{Erickson:2021psj}, gravity localisation is characterised by a calculable Newton constant and a clear transition between higher-dimensional near-field behaviour and lower worldvolume-dimensional far-field behavior as the spatial worldvolume separation between gravitating sources increases. Gravity localisation can take place with or without a mass gap -- the key point is the existence of a normalisable transverse wavefunction zero mode. The corresponding difference in the effective theories is between Yukawa versus power-law corrections to the long-distance lower-dimensional behaviour, as has been clear since the time of the RSII construction \cite{Randall:1999vf}.

So far, this investigation has been carried out just in a classical or semiclassical supergravity context and a next level should be to see how much of this structure is realised at the quantum/superstring level.

From the localising examples found, we observe that only the transverse spaces that are asymptotically $\mb{R}^2\times\{\text{compact}\}$ at small distances\footnote{The origin of the $\mb{R}^2$ is the brane worldvolume.} admit a normalisable zero mode (except for RSII, which does not have a resolved brane structure). The pattern is that the measure $\mu(z)$ is always a power of $z$ near $z=0$ where the zero eigenfunction is logarithmically divergent. It is the logarithmic behaviour of the eigenfunction that allows for its normalisability, even though the transverse space is non-compact. As such, we conjecture that only resolved branes with transverse spaces that asymptote to $\mb{R}^2\times\{\text{compact}\}$ at small distances allow for worldvolume gravity localisation.

\section*{Acknowledgments}
We are grateful to Costas Bachas and Chris Pope for helpful discussions. The work of KSS was supported in part by the STFC under Consolidated Grants ST/T000791/1 and ST/X000575/1.

\section*{Appendices} 
\addcontentsline{toc}{section}{Appendices}

\begin{appendix}

\section{M-theory and Type IIA supergravity conventions} \label{Mtheory}

The bosonic sector of M-theory consists of a metric $\hat g_{MN}$ and a four-form flux $\hat F_{(4)}$. The Bianchi identity for the four-form is
\begin{equation}\label{bianchiF4}
d\hat F_{(4)} = 0 \,,
\end{equation}
and the equations of motion are
\begin{eqnarray}
&&d\hat{\ast}\hat F_{(4)} - \frac{1}{2}\hat F_{(4)}\wedge\hat F_{(4)} = 0 \,, \label{eomF411d} \\
&&\hat R_{MN} = \frac{1}{12}\left(\hat F_{MPQR}\hat F_N^{\ph{N}PQR} - \frac{1}{12}(\hat F_{(4)})^2\hat g_{MN}\right) \,.\label{11einstein}
\end{eqnarray}
The dynamics of Type IIA supergravity are encoded in M-theory about backgrounds where the eleven-dimensional spacetime has a $U(1)$ isometry. We write
\begin{equation}
\begin{split}
&d\hat s^2_{11} = e^{-\tilde\Phi/6}d\tilde s^2_{10}  +e^{4\tilde\Phi/3}(d\theta + \tilde A_{(1)})^2 \,,\\
&\hat F_{(4)} = \tilde F_{(4)} + \tilde H_{(3)}\wedge (d\theta + \tilde A_{(1)}) \,,
\end{split}
\end{equation}
where the ten-dimensional Type IIA fields are distinguished by a tilde and are independent of $\theta$. The metric $\tilde g$ is the Einstein frame metric of the Type IIA theory. The Bianchi identity of $\hat F_{(4)}$ yields
\begin{equation}
d\tilde H_{(3)} = 0 \,,\quad d\tilde F_{(4)} - \tilde H_{(3)}\wedge \tilde F_{(2)} = 0 \,,
\end{equation}
where $\tilde F_{(2)} = d\tilde A_{(1)}$. These can be integrated by introducing the NSNS two-form potential $\tilde B_{(2)}$ and the RR three-form potential $\tilde A_{(3)}$,
\begin{equation}
\tilde H_{(3)} = d\tilde B_{(2)} \,,\quad \tilde F_{(4)} = d\tilde A_{(3)} - \tilde H_{(3)}\wedge \tilde A_{(1)} \,.
\end{equation}

\section{Type IIB supergravity conventions} \label{IIBeoms}

Our conventions are taken from \cite{Gauntlett:2010vu}. The bosonic sector of Type IIB supergravity contains the RR forms $\hat F_{(1)}$, $\hat F_{(3)}$, $\hat F_{(5)}$, the NSNS three-form $\hat H_{(3)}$, the dilaton $\hat\Phi$ and the metric, which will be taken to be the Einstein frame metric. The Bianchi identities are 
\begin{eqnarray}
&& d\hat F_{(5)} + \hat F_{(3)} \wedge  \hat H_{(3)} =0 \label{F5} \,,\\
&& d\hat F_{(3)}  + \hat F_{(1)} \wedge \hat H_{(3)}=0 \label{F3}\,, \\
&& d\hat H_{(3)} =0 \label{H3}\,,\\
&& d\hat F_{(1)} =0 \label{F1} \,,
\end{eqnarray}
which can be solved by introducing potentials:
\begin{equation}
\begin{split}
&\hat F_{(1)} = d\hat C_0 \,,\quad \hat H_{(3)} = d\hat B_{(2)}\,,\\
&\hat F_{(3)} = d\hat C_{(2)} -\hat C_0  d\hat B_{(2)}\,,\quad \hat F_{(5)} = d\hat C_{(4)} -\hat C_{(2)} \wedge  \hat H_{(3)} \,.
\end{split}
\end{equation}
The equations of motion read
\begin{eqnarray}
&& \hat{\ast}\hat F_{(5)} =\hat F_{(5)} \label{eomF5} \,,\\
&& d(e^{\hat\Phi} \hat{\ast}\hat F_{(3)})  - \hat F_{(5)} \wedge \hat H_{(3)}=0 \label{eomF3}\,, \\
&& d(e^{-\hat\Phi} \hat{\ast}\hat H_{(3)})  -e^{\hat\Phi} \hat F_{(1)} \wedge \hat{\ast}\hat F_{(3)} -\hat F_{(3)} \wedge \hat F_{(5)}  =0 \label{eomH3} \,,\\
&& d(e^{2\hat\Phi} \hat{\ast}\hat F_{(1)})  +e^{\hat\Phi} \hat H_{(3)} \wedge \hat{\ast}\hat F_{(3)} =0 \label{eomF1} \,,\\
&& d\hat{\ast}d\hat\Phi -e^{2\hat\Phi} \hat F_{(1)} \wedge \hat{\ast}\hat F_{(1)} +\frac{1}{2}e^{-\hat\Phi} \hat H_{(3)} \wedge \hat{\ast}\hat H_{(3)} -\frac{1}{2} e^{\hat\Phi} \hat F_{(3)} \wedge \hat{\ast}\hat F_{(3)} =0 \label{eomPhi}\,,\\
&&\hat R_{MN} =  \frac{1}{2} \hat\nabla_M \hat\Phi \hat\nabla_N \hat\Phi+\frac{1}{2} e^{2\hat\Phi} \hat\nabla_M C_{0} \hat\nabla_N C_{0}
      + \frac{1}{96}\hat F^{}_{MP_1P_2P_3P_4}\hat F_N^{\ph{M}P_1P_2P_3P_4}  \nonumber \\ && \quad \qquad
      + \frac{1}{4}e^{-\hat\Phi}\left(
         \hat H^{}_{MP_1P_2} \hat H_{N}^{\ph{N}P_1P_2}
         - \frac{1}{12}(\hat H_{(3)})^2\hat g_{MN} \right) \nonumber \\ && \quad \qquad
      + \frac{1}{4}e^{\hat\Phi}\left(
         \hat F^{}_{MP_1P_2} \hat F_{N}^{\ph{N}P_1P_2}
         - \frac{1}{12}(\hat F_{(3)})^2\hat g_{MN}\right) \,. \label{IIBEinstein}
\end{eqnarray}
These admit a trombone symmetry, where 
\begin{equation}\label{tromboneIIB}
\begin{split}
&\hat g_{MN}\mapsto k^2 \hat g_{MN}\,,\quad \hat\Phi\mapsto \hat\Phi\,,\quad \hat C_0\mapsto \hat C_0\,,\\
& \hat F_{(3)}\mapsto k^2\hat F_{(3)}\,,\quad \hat H_{(3)}\mapsto k^2\hat H_{(3)}\,,\quad \hat F_{(5)}\mapsto k^4\hat F_{(5)} \,,
\end{split}
\end{equation}
for a constant $k$. 

In a configuration where $\hat\Phi = 0$ and $\hat F_{(1)} = 0$, it is convenient to package the three-form fluxes into a complex three-form
\begin{equation}
\hat G_{(3)} = \hat F_{(3)} + i\hat H_{(3)} \,,
\end{equation}
whose Bianchi identity and equation of motion are
\begin{equation}
d\hat G_{(3)} = 0 \,,
\end{equation}
and
\begin{equation}\label{eomG3}
d\hat {\ast}\hat G_{(3)} - i\hat G_{(3)}\wedge \hat F_{(5)} = 0 \,.
\end{equation}
The Bianchi identity for the five-form then reads
\begin{equation}\label{eomF5G3}
d\hat F_{(5)} + \frac{i}{2}\hat G_{(3)}\wedge\ov{\hat{G}}_{(3)} = 0 \,,
\end{equation}
and the consistency conditions for setting $\hat\Phi= 0$ and $\hat F_{(1)} = 0$ are contained in the single condition
\begin{equation}\label{compatibilityG3}
\hat G_{(3)}\wedge\hat{\ast}\hat G_{(3)} = 0 \,.
\end{equation}
With this condition, the Einstein equations simplify to
\begin{equation}
\hat R_{MN} = \frac{1}{96}\hat F^{}_{MP_1P_2P_3P_4}\hat F_N^{\ph{M}P_1P_2P_3P_4} + \frac{1}{8}\left(\hat G^{}_{MP_1P_2}\ov{\hat G}_N^{\ph{N}P_1P_2} + \ov {\hat G}^{}_{MP_1P_2} \hat G_N^{\ph{N}P_1P_2}\right) \,.
\end{equation}

\section{Bucher rules for Abelian T-duality} \label{Bucher}

We provide here a brief review of the Buscher rules for Abelian T-duality, following the presentation in Ref.\ \cite{Kelekci:2014ima}. The rules for the NSNS sector were derived in \cite{Buscher:1985kb,Buscher:1987qj}, and those for the RR sector were derived in \cite{Hassan:1999bv}. 

Suppose we have a (massive) Type IIA background with a $U(1)$ isometry generated by $\partial/\partial\psi$. The most general solution compatible with this symmetry is given by
\begin{equation}
\begin{split}
&d\tilde s^2_{str} = ds^2_9 + e^{2A}(d\psi + A_{(1)})^2 \,, \\
&\tilde F_{(0)} = m \,,\\
&\tilde B_{(2)} = L_{(2)} + B_{(1)}\wedge d\psi\,,\\
&\tilde F_{(2)} = G_{(2)} + G_{(1)}\wedge(d\psi+A_{(1)}) \,,\\
&\tilde F_{(4)} = G_{(4)} + G_{(3)}\wedge(d\psi + A_{(1)}) \,,
\end{split}
\end{equation}
where $A$, $ds^2_9, L_{(2)}, B_{(1)}$, $G_{(i)}$, and the dilaton $\tilde\phi$ are independent of $\psi$, where $m$ is the Romans mass, and the metric is in string frame. The assumption that the NSNS 2-form $\tilde B_{(2)}$ is independent of $\psi$ can be relaxed slightly by noting that only its field strength $\tilde H_{(3)}$ appears in the equations of motion even though $\tilde B_{(2)}$ itself appears in the string sigma model. As long as $\tilde H_{(3)}$ is of the form $L_{(3)} + H_{(2)}\wedge(d\psi + A_{(1)})$, one can T-dualise along $\psi$, since $\partial/\partial\psi$ will remain a Killing vector. There are generalisations required in the Buscher rules for such cases, but these will not be necessary here, as such solutions are not our main focus. 

T-dualising along $\psi$ gives a Type IIB solution,
\begin{equation}
\begin{split}
&d\hat s^2_{str} = ds^2_9 + e^{-2A}(d\xi + B_{(1)})^2 \,, \\
&e^{\hat\Phi} = e^{\phi-A} \,, \\
&\hat F_{(1)} = -G_{(1)} + m(d\xi + B_{(1)}) \,,\\
&\hat B_{(2)} = L_{(2)} + A_{(1)}\wedge(d\xi + B_{(1)}) \,,\\
&\hat F_{(3)} = -G_{(3)} + G_{(2)}\wedge(d\xi +B_{(1)}) \,,\\
&\hat F_{(5)} = (1+\hat{\ast})G_{(4)}\wedge(d\xi+B_{(1)}) \,,
\end{split}
\end{equation}
where $\xi$ is the dual coordinate of $\psi$, the metric is in string frame, and we set $\alpha' = 1$.

\section{Type IIB embedding of the Salam-Sezgin theory} \label{Salam-Sezgin T-dual}

The Salam-Sezgin theory is an $\mc{N}=2$ six-dimensional supergravity whose bosonic sector contains a metric, a two-form gauge potential, a vector, and a scalar \cite{Salam:1984cj}. The Lagrangian for the bosonic sector is
\begin{equation}
\mc{L}_6 = \ov R\ov{\ast}_61 - \frac{1}{4}d\ov\phi\wedge\ov{\ast}_6d\ov\phi - \frac{1}{2}e^{-\ov\phi/2}\ov F_{(2)}\wedge\ov{\ast}_6\ov F_{(2)} - \frac{1}{2}e^{\ov\phi}\ov H_{(3)}\wedge\ov{\ast}\ov H_{(3)} - 8\ov g^2e^{-\ov\phi/2}\ov{\ast}_61\,,
\end{equation}
where $\ov F_{(2)} = d\ov A_{(1)}$, $\ov H_{(3)} = d\ov B_{(2)} - \tfrac{1}{2}\ov A_{(1)}\wedge\ov F_{(2)}$, and we put bars over six-dimensional fields. In \cite{Cvetic:2003xr}, this theory was embedded into the NSNS sector of Type IIA supergravity with
\begin{equation}
\begin{split}
&d\tilde s^2 = (\cosh2\rho)^{1/4}\left(e^{-\ov\phi/4}d\ov s^2_6 + e^{\ov\phi/4}dy^2 + \frac{1}{2\ov g^2}e^{\ov\phi/4}\left[d\rho^2 + \frac{1}{4}\tanh^22\rho\left(d\chi -2\ov g\ov A_{(1)}\right)^2 \right.\right. \\
&\left.\left.\qquad\qquad\qquad\qquad+ \frac{1}{4}\left(d\psi + \sech2\rho\,(d\chi -2\ov g\ov A_{(1)})\right)^2\right]\right) \,,\\
&\tilde H_{(3)} = \ov H_{(3)} - \frac{\sinh2\rho}{4\ov g^2\cosh^22\rho}d\rho\wedge d\psi\wedge (d\chi -2\ov g\ov A_{(1)}) \\
&\qquad\quad+ \frac{1}{4\ov g\cosh2\rho}\ov F_{(2)}\wedge\left(d\psi + \cosh2\rho\,(d\chi -2\ov g\ov A_{(1)})\right) \,,\\
&e^{\tilde\phi} = (\cosh2\rho)^{-1/2}e^{-\ov\phi/2}\,,
\end{split}
\end{equation}
where the metric is in Einstein frame. We will now lift this into M-theory on a circle parametrised by $\zeta$ and then reduce the solution back to Type IIA on the circle parametrised by $y$. The resulting solution in string frame is 
\begin{align*}
&d\tilde s^2_{str} = (\cosh2\rho)^{1/2}\left(d\ov s^2_6 + e^{\ov\phi/2}dy^2 + \frac{1}{2\ov g^2}e^{\ov\phi/2}\left[d\rho^2 + \frac{1}{4}\tanh^22\rho\left(d\chi -2\ov g\ov A_{(1)}\right)^2+ \left(d\psi + \mc{A}_{(1)}\right)^2\right]\right) \\
&\qquad\quad+ (\cosh2\rho)^{-1/2}e^{-\ov\phi/2}d\zeta^2 \,,\\
&\tilde F_{(4)} = \left(\ov H_{(3)} + \frac{1}{4\ov g}\tanh^22\rho\,\ov F_{(2)} \wedge (d\chi-2\ov g\ov A_{(1)}) \right)\wedge d\zeta \numberthis \\
&\qquad\quad + \left(\frac{1}{\ov g^2}\tanh2\rho\,d\rho\wedge d\zeta\wedge \mc{A}_{(1)} - \frac{1}{2\ov g}\sech2\rho\,\ov F_{(2)}\wedge d\zeta\right)\wedge(d\psi + \mc{A}_{(1)}) \,,\\
&e^{\tilde\phi} = (\cosh2\rho)^{1/4}e^{\ov\phi/4}\,.
\end{align*}
where we have defined
\begin{equation}
\mc{A}_{(1)} = \frac{1}{2}\sech2\rho \,(d\chi -2\ov g\ov A_{(1)}) \,,
\end{equation}
and rescaled $\psi\mapsto\psi/2$. Applying the Bucher rules presented in Appendix \ref{Bucher} on the coordinate $\psi$, we obtain the Type IIB solution
\begin{equation}\label{IIBSS}
\begin{split}
&d\hat s^2_{str} = (\cosh2\rho)^{1/2}\left(d\ov s^2_6 + \frac{1}{2\ov g^2}e^{\ov\phi/2}\left[d\rho^2 + \frac{1}{4}\tanh^22\rho\left(d\chi -2\ov g\ov A_{(1)}\right)^2\right]\right) \\
&\qquad\quad+ (\cosh2\rho)^{-1/2}e^{-\ov\phi/2}\left(d\zeta^2 + 2\ov g^2 d\xi^2\right) \,,\\
&e^{\hat\Phi} = \sqrt{2}\ov g\,,\quad \hat C_{0} = 0 \,,\\
&\hat H_{(3)} = -d\mc{A}_{(1)}\wedge d\xi\,,\quad \hat F_{(3)} = \frac{1}{2\ov g^2}d\mc{A}_{(1)}\wedge d\zeta\,, \\
&\hat F_{(5)} = (1+\hat{\ast})\left[\left(\ov H_{(3)} + \frac{1}{4\ov g}\tanh^22\rho\,\ov F_{(2)} \wedge (d\chi-2\ov g\ov A_{(1)}) \right)\wedge d\zeta \wedge d\xi\right] \,,
\end{split}
\end{equation}
where we have flipped the signs $(\zeta,\xi)\mapsto (-\zeta,-\xi)$. We can freely shift the dilaton vacuum value to zero by an $SL(2,\mb{R})$ transformation with matrix $\Lambda = \diag(p,1/p)$, where $p = 2^{1/4}\ov g^{1/2}$, which also rescales $\hat F_{(3)}$ and $\hat H_{(3)}$ by $p$ and $1/p$ respectively. 

We can interpret \eqref{IIBSS} as a truncation of the Type IIB theory on a four-manifold $M_2\times T^2$, where $T^2$ is the two-torus parametrised by $(\zeta,\xi)$, and $M_2$ is a non-compact two-manifold with metric
\begin{equation}
ds^2_2 = \cosh2\rho\,d\rho^2 + \frac{1}{4}\frac{\sinh^22\rho}{\cosh2\rho}d\chi^2 \,.
\end{equation}
From \eqref{EHmetric}, we recognise this two-dimensional metric to be the induced metric on Eguchi-Hanson space where the two-sphere coordinates are held constant; $ds^2_2 = ds^2_{EH}\rvert_{(\theta,\varphi)=(\theta_0,\varphi_0)}$. As $\rho\to0$, the metric becomes
\begin{equation}
ds^2_2\big\rvert_{\rho\to0} = d\rho^2 + \rho^2d\chi^2 \,.
\end{equation}
The coordinate $\chi \in [0,2\pi)$, so $M_2$ is $\mb{R}^2$ in this limit. For the asymptote at infinity, it is useful to consider a change of coordinates $r^2 = \cosh2\rho$, in which the metric becomes
\begin{equation}
ds^2_2 = \left(1-\frac{1}{r^4}\right)^{-1}dr^2 + \frac{r^2}{4}\left(1-\frac{1}{r^4}\right)d\chi^2 \,.
\end{equation}
From this, we find that as $r\to\infty$,
\begin{equation}
ds^2_2\big\rvert_{r\to\infty} = dr^2 + \frac{r^2}{4}d\chi^2 \,,
\end{equation}
so $M_2$ is $\mb{R}^2$ with a conical deficit angle $\pi$ in this limit, which is simply the upper half-plane $\mb{R}\times\mb{R}^+$. The Euler characteristic of $M_2$ can be computed using the Gauss-Bonnet theorem, 
\begin{equation}
\chi(M_2) = \frac{1}{2\pi}\left(\int_{M_2} K\vol(M_2) + \int_{\partial M_2}\kappa\, ds \right) = 1\,.
\end{equation}
where $K = R/2$ is the Gaussian curvature, $\partial M_2$ is the boundary of $M_2$ at infinity with line element $ds$, and $\kappa$ is the geodesic curvature at infinity. Note however that the Euler characteristic of the four-manifold $M_2\times T^2$ is 
\begin{equation}
\chi(M_2\times T^2) = \chi(M_2)\chi(T^2) = 0 \,.
\end{equation}
The two-manifold $M_2$ is K\"ahler. It is easiest to see this by performing the change of coordinates $u^4 = r^4-1$, after which the metric reads
\begin{equation}
ds^2_2 = \left(1+\frac{1}{u^4}\right)^{-1/2}\left(du^2 + \frac{u^2}{4}d\chi^2\right) \,.
\end{equation}
Defining the complex coordinate $w = ue^{i\chi/2}$, we can rewrite the metric as
\begin{equation}
ds^2_2 = \left(1+\frac{1}{|w|^4}\right)^{-1/2}dwd\ov w \,,
\end{equation}
from which we deduce the K\"ahler potential
\begin{equation}
K = (1 + |w|^4)^{1/2} - \tanh^{-1}\left((1 + |w|^4)^{1/2}\right) \,.
\end{equation}
Since the two-torus is also K\"ahler, the four-manifold $M_2\times T^2$ is K\"ahler.

\addcontentsline{toc}{section}{References}

\bibliographystyle{utphys}
\bibliography{RBW}{}

\end{appendix}

\end{document}